\pdfoutput=1
\documentclass{sig-alternate-05-2015}

\usepackage{graphicx}
\usepackage{subfig}
\usepackage{subfloat}
\usepackage{amsmath}
\usepackage{amsfonts}
\usepackage{multirow}
\usepackage{url}

\begin{document}
\setcopyright{none}

\doi{}

\isbn{}

\title{Modeling Dense Urban Wireless Networks with 3D Stochastic Geometry
}


\author{
\alignauthor
Alexandre Mouradian\\
       \affaddr{Laboratoire des Signaux et Syst\`emes (L2S, UMR8506),}\\
       \affaddr{Univ. Paris Sud-CNRS-CentraleSup\'elec,}
       \affaddr{Universit\'e Paris Saclay}\\
       \affaddr{F-91192, Gif-sur-Yvette}\\
       \email{alexandre.mouradian@u-psud.fr}
}


\maketitle

\begin{abstract}
Over the past decade, many works on the modeling of wireless networks using stochastic geometry have been proposed. Results about probability of coverage, throughput or mean interference, have been provided for a wide variety of networks (cellular, ad-hoc, cognitive, sensors, etc). These results notably allow to tune network protocol parameters. Nevertheless, in their vast majority, these works assume that the wireless network deployment is flat: nodes are placed on the Euclidean plane. However, this assumption is disproved in dense urban environments where many nodes are deployed in high buildings.
In this paper, we derive the exact form of the probability of coverage for the cases where the interferers form a 3D Poisson Point Process (PPP) and an approximation for the 3D Modified Matern Process (MMP). We compare the 3D model with the 2D model and with simulation results. We comment the adequacy of each model in function of the parameters of the nodes (emission power, reception threshold, MAC protocol, etc.) and the height of the buildings in the simulations.
\end{abstract}


\section{Introduction}
\label{introsec}

Stochastic geometry have been largely used to study and design wireless networks, because in such networks the interference and thus the capacity is highly dependent on the positions of the nodes \cite{baccelli09,haenggi09}. Stochastic geometry indeed allows to take into account the spatial component for the analysis of wireless systems performance at a very low computational cost (in several cases closed form expressions are available \cite{baccelli09}). Nevertheless, most of the works in the literature focus on networks deployed on the Euclidean plane. The motivation for this work comes from the intuition that for dense urban networks such as WiFi private networks, sensor networks, connected objects  deployed in dense urban areas, modeling the network by projecting all the nodes on a plane will lead to an inaccurate representation of the interference suffered by a node. We think that with the explosion of the number of devices expected in the ISM bands, notably because of smartphones, tablets, connected objects (Internet of Things) and sensors, there is a need for an accurate modeling of the interference suffered by such devices in dense urban areas. Indeed, accurate interference modeling is essential in order to design efficient mitigation mechanisms at different layers (medium access schemes, coding schemes, retransmission mechanisms, robust routing, etc.). In order to be accurate, the third space dimension has to be taken into account.

Throughout this paper, we model the node positions as a Poisson Point Process (PPP). We argue that the PPP seems to be a good model for dense urban areas where many networks are concurrently deployed in the same frequency bands (notably ISM bands) in a chaotic manner. Unlike cellular networks, for WiFi, sensor and IoT networks no planned deployment can be assumed in general \cite{akella07}.

For the numerical applications, we take the example of private WiFi networks in a dense urban area such as the center of Paris. We consider, the \emph{VIe arrondissement} of Paris which contains around 32500 dwellings \cite{paris} and covers a surface of 2.15km$^2$. We assume that each dwelling has one WiFi router. In the 2D case, they form a Poisson Point Process (PPP) on the plane. In the 3D case, they are also spread on the $z$ axis. If we assume a mean building height of 20 meters for the 3D case, it gives intensities of $\lambda=1.51\times 10^{-2}$ nodes per $m^2$ and $\rho=7.56\times 10^{-4}$ nodes per $m^3$ respectively for the 2D and 3D processes. These intensities may seem high, but they actually do not take into account the devices (laptops and smartphones for instance) connected to the routers which also increase the interference.

In the remainder of this paper, we derive the probability of coverage, first with no medium access control (the results are applicable to ALOHA \cite{baccelli09}), and then with a CSMA access when nodes are distributed according to a 3D PPP.

The main contributions of this paper are as follows:
\begin{itemize}
 \item we give an approximation of the probability of coverage for CSMA access in the 3D case;
 \item we compare 2D, 3D analytical models and simulation in order to determine the most suited model in function of the parameters of nodes and network topology (notably the height of the deployment along the $z$ axis).
\end{itemize}

The paper is organized as follows. Section \ref{RW} presents and comments related works on stochastic geometry and 3D network modeling. In Section \ref{PPPsec}, we derive the exact form of the probability of coverage for PPP interference for arbitrary emitter-receiver distance and compare the results with the 2D model and simulations. It gives us insights on the problem for a simple case. In Section \ref{MMPsec}, we present the Modified Matern Process used to model CSMA access and derive an approximation of the probability of coverage for the 3D case. We then compare this approximation with the one of the 2D case and simulations. Section \ref{conclusec} concludes the discussion and provides planned future developments.

\section{Related works}
\label{RW}

Even if most of the research efforts are focused on 2D networks, 3D networks have been emerging during the past few years. It is notably the case for WSNs \cite{zhou10,ammari10,doddavenkatappa12} underwater mobile networks \cite{cui06} Unmanned Aerial Vehicle (UAV) networks \cite{valente11}.

From a more theoretical point of view, 3D networks have been investigated in terms of capacity \cite{gupta00,li11} and scaling laws have been provided. Nevertheless, in the present work we focus on the stochastic geometric approach to the study of wireless networks in the sense of \cite{baccelli09,haenggi09}. This approach has the advantage to provide tractable results for the probability of coverage in the case of the PPP model and good approximations for other models \cite{baccelli09}.

Whereas, many theoretical works focusing on stochastic geometry for wireless networks consider dimension $d$ \cite{baccelli09,haenggi09}, when it comes to applications to specific cases, the chosen space is nearly always the Euclidean plane \cite{nguyen07,baccelli09,elsawy13}. The only work explicitly covering the 3D case, to the best of our knowledge, is the recent \cite{gupta15}. In this work, the authors derive the probability of coverage with the two slopes propagation model for a 3D PPP. This work considers cellular networks, the receiver is thus attached to the nearest point of the PPP (the base station with the highest average received power). In our work, we first focus on a 3D PPP model as well, but with a single slope propagation model and, more importantly, the receiver is receiving from an arbitrarily placed emitter. We argue that this model is more relevant in non operated ISM band networks such as WiFi, IoT, WSNs, etc, because the receiver cannot always connect to the closest node. Indeed, the closest node might not belong to the same network. Moreover, in our work, we mainly use the simple PPP model to easily compare simulation and theoretical models for different building heights (maximum value on the $z$ axis) and thus evaluating the relevance of the 3D approach. We also then move to the Modified Matern Process which is a more realistic model CSMA networks.

Many works modeling CSMA access through stochastic geometry have emerged in the literature these past ten years \cite{nguyen07,kaynia11,elsawy13}. These works are based on a modified version of the Matern Point Process for which the points cannot live too close to each others. This allows to model the contention radius of CSMA as follows. The nodes of the network form a PPP on the Euclidean plane and they contend to access to the medium. Each node picks uniformly a value between zero and one, the node is selected in the process if it does not detect any node with a smaller mark value (this models the backoff procedure). The resulting process is the Modified Matern Process (MMP) which will be detailed in section \ref{MMPsec}. As there is no known exact formulation of the interference in such process, \cite{nguyen07,kaynia11} and \cite{elsawy13} use approximation techniques. They show that their approximations lie close to simulations results. We extend these works by considering 3D distribution of the nodes and compare the resulting model to simulation. To the best of our knowledge, this has not been considered in the literature.

In the remainder of this paper we show how the probability of coverage is changed when going from 2D to 3D for Poisson distributed nodes in the simple case where all nodes can be interferers, and in the case a CSMA access protocol is used. We compare the theoretical prediction with simulations. We also show that going from 2D to 3D is not trivial especially in the CSMA case because changing the dimension dramatically affects the form and thus the tractability of the expressions.

\section{3D Poisson Point Process interference}
\label{PPPsec}
In this section we derive the probability of coverage for a terminal at distance $d$ from the emitter when the interferers are distributed according to a PPP $\Phi$ in $\mathbb{R}^3$. Whereas it is already given for $\mathbb{R}^n$ and for $\mathbb{R}^2$ in \cite{baccelli09}, here we find interesting to emphasis the differences between the 3D and 2D cases and to show that considering the latter can lead to inaccurate representation of the interference.
We consider an interference limited network. We thus use the Signal to Interference Ratio (SIR), which is defined as follows:

\begin{equation}
 SIR=\frac{hd^{-\alpha}}{I}
 \label{sireq}
\end{equation}

with $h$ the fading coefficient between the emitter and receiver, $d$ the emitter-receiver distance, $\alpha$ the pathloss exponent and $I=\sum_{i\in \Phi}g_ir_i^{-\alpha}$ the interference where $g_i$ is the fading coefficient between the interferer $x_i \in \Phi$ and the receiver and $r_i$ the distance between them. As in \cite{nguyen07,hasan07}, we assume that all the nodes emit with the same power so it is simplified in the expression of the SIR. We have to note that unlike many works in the literature, we do not consider that the receiver is connected to the nearest node of the PPP because it is not always the case for the type of networks we consider (private WiFi is a good example) so the probability of coverage depends on the emitter-receiver distance. The probability of coverage is defined as the probability that the SIR is over a given threshold $\beta$:
\begin{equation}
 P_c(\rho,\beta,\alpha,d)\equiv P\{SIR>\beta\}
\end{equation}
Its expression, when considering Rayleigh fading between the emitter and receiver, can be derived by the classic argument notably found in \cite{baccelli09} and is equal to the Laplace transform of the interference shot noise in interference limited networks:

\begin{align}
 P_c(\rho,\beta,\alpha,d) & = P\left(\frac{hd^{-\alpha}}{I}>\beta\right) \nonumber\\
 & \stackrel{(a)}{=} \mathbb{E}_{I}[e^{-\mu d^{\alpha}\beta I}] \nonumber\\
 & = \mathcal{L}_I(\mu \beta d^{\alpha})
\end{align}

(a) follows from $H\sim exp(\mu)$.

The expression of the Laplace transform for the 3D PPP case is as follows:

\begin{align}
 \label{laplaceeq}
&\mathcal{L}_I(\mu \beta d^{\alpha}) \nonumber\\
 & \hphantom{aaa} = \mathbb{E}_{G_i,R_i}[e^{-\mu d^{\alpha}\beta \sum \limits_{x_i \in \Phi} g_ir_i^{-\alpha} }] \nonumber\\
 & \hphantom{aaa}\stackrel{(b)}{=} \mathbb{E}_{R_i}[\prod \limits_{x_i \in \Phi} \mathbb{E}_{G_i}[e^{-\mu d^{\alpha}\beta g_i r_i^{-\alpha} }]]  \nonumber\\
 & \hphantom{aaa}\stackrel{(c)}{=} e^{-\rho \int \limits_{0}^{+\infty} \int \limits_{0}^{2\pi} \int \limits_{0}^{\pi} (1-\mathbb{E}_{G_i}[e^{-\mu d^{\alpha}\beta g_i r_i^{-\alpha} }])r_i^2sin\theta_i dr_i d\theta_i d\phi_i} \nonumber\\
 & \hphantom{aaa}\stackrel{(d)}{=} e^{-\rho 4 \pi \int \limits_{0}^{+\infty}(1-\frac{1}{1+\beta d^{\alpha} r_i^{-\alpha}})r_i^2 dr_i} \nonumber\\
 & \hphantom{aaa}\stackrel{(e)}{=} e^{-\rho 4 \pi \frac{d^3}{3} \beta^{3/\alpha} \int \limits_{0}^{+\infty}(\frac{1}{1+u^{\alpha/3}})du}
 \phantom{\hspace{2cm}}
\end{align}

(b) follows from $G_i$ being i.i.d. and independence with $R_i$, (c) follows from the probability generating functional of the PPP \cite{stoyan87}, (d) follows from the MGF of $G_i \sim exp(\mu)$ and (e) with the change of variable $u=\left(\frac{r_i}{d\beta^{1/\alpha}} \right)^3$. It differs from the 2D case which is given by $ e^{-\lambda \pi d^2 \beta^{2/\alpha} \int \limits_{0}^{+\infty}(\frac{1}{1+v^{\alpha/2}})dv}$. We note that the solution of the integral in the general case (dimension $D$) is:
\begin{equation}
  \int \limits_{0}^{+\infty}(\frac{1}{1+s^{\alpha/D}})ds=s\times {}_2F_1(1,\frac{D}{a},\frac{D}{a}+1,-s^{a/d})
\end{equation}

with ${}_2F_1$ the Gaussian hypergeometric function. This function gives highly different values even for the same pathloss exponent $\alpha$ in the 2D and the 3D cases. The other main difference is the exponential decay which is in $d^2$ in the 2D case and $d^3$ in 3D. We give the closed form for $\alpha=4$ for the 3D case:

\begin{equation}
 \label{pc3Dpppeq}
 P_c(\rho,\beta,4,d)=e^{-\rho (4/2^{3/2})  \pi^2 d^3 \beta^{3/4}}
\end{equation}

\begin{figure}[ht]
  \centering
  \includegraphics[width=3in, keepaspectratio=true]{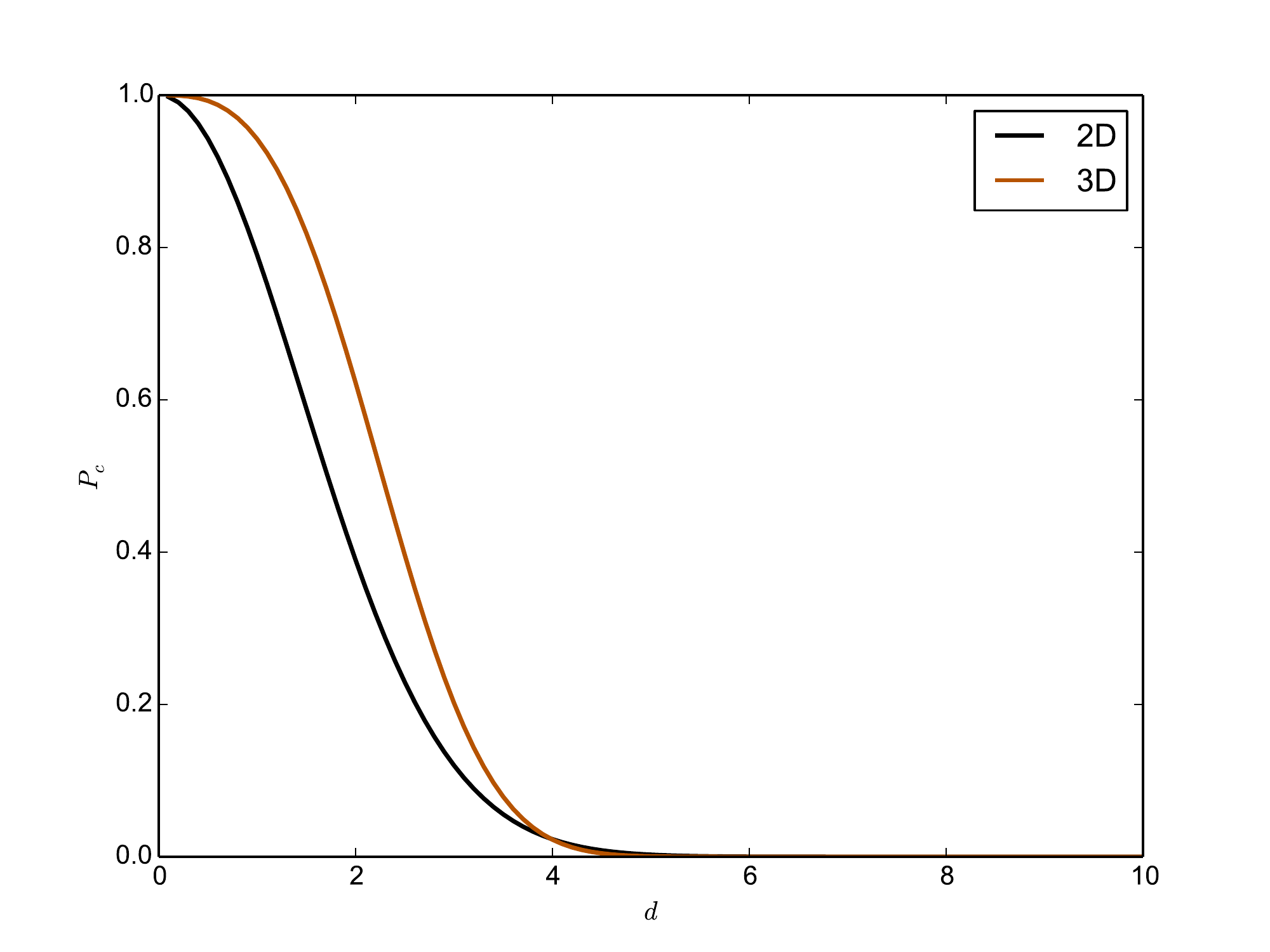}
  \caption{Probability of coverage for 2D and 3D PPP interference}
  \label{pppfig}
\end{figure}

In Fig. \ref{pppfig}, we plot $P_c(\rho,\beta,4,d)$ for the 2D and 3D cases with $\lambda=1.51\times 10^{-2}$ and $\rho=7.56\times 10^{-4}$ (as motivated in Section \ref{introsec}) and $\beta=10$, the plot is in function of the emitter-receiver distance $d$. We observe that $P_c$, in the 2D case, is highly underestimated when the receiver is close to the emitter. Given the steepness of the decay, the error might be important: for instance, there is 0.23 absolute difference at 2 meters. We note that the $P_c$ is getting very rapidly low (after few meters $P_c$ goes to zero) which is not realistic. We observe this because, in this simple case, all the nodes transmit at the same time, whereas it is not the case in reality where a MAC protocol is used. 
In Section \ref{MMPsec}, we focus on the modeling of a more realistic setup: a 3D CSMA network. Nevertheless, We can note that expression (\ref{pc3Dpppeq}) is valid for modeling an ALOHA medium access. In this case, the intensity of the process should be multiplied by the probability for a node to emit \cite{baccelli09}. ALOHA access is currently one of the considered access mechanism in the IoT standard LoRaWAN \cite{lorawan} from the LoRa Alliance \cite{loraaliance}. The 3D PPP model could thus be used to model such IoT networks in dense urban areas. In the next subsection, we present simulation results which advocate for the relevancy of the 3D model.

\subsection{Comparison with simulation results}
\label{pppsimsec}

In this section, we compare the theoretical and simulation results. We evaluate the relevancy of the 3D approach for dense urban areas by considering the following question: for which building height does it makes sense to model the interference process as a 3D PPP? Indeed, in reality, the nodes will be spread within a finite height because of the limited height of the buildings, so a border effect may appear compared to the theoretical model.

The simulation setup is as follows: the point process is generated in box of $200m\times 200m \times Z$ with $Z\in {10m,50m,100m}$. The 3D intensity of the process is calculated based on the 2D intensity: $\rho=\frac{\lambda}{Z}$. The reason for this is that in the 2D case we project all the points on a plane, the intensity is thus maximal. In the 3D case the points are also spread along the $z$ axis. Then for each realization of the PPP, the SIR is computed at the origin (the center of the aforementioned box) thanks to equation (\ref{sireq}). The simulation and model parameters are the following: $\lambda=1.51\times 10^{-2}$ (we keep the values from the example of Section \ref{introsec}), $\beta=10$ and $\alpha=4$. The emitter-receiver distance $d$ varies from $0$ to $10$ meters. 

\begin{figure}[ht]
  \centering
  \includegraphics[width=3in, keepaspectratio=true]{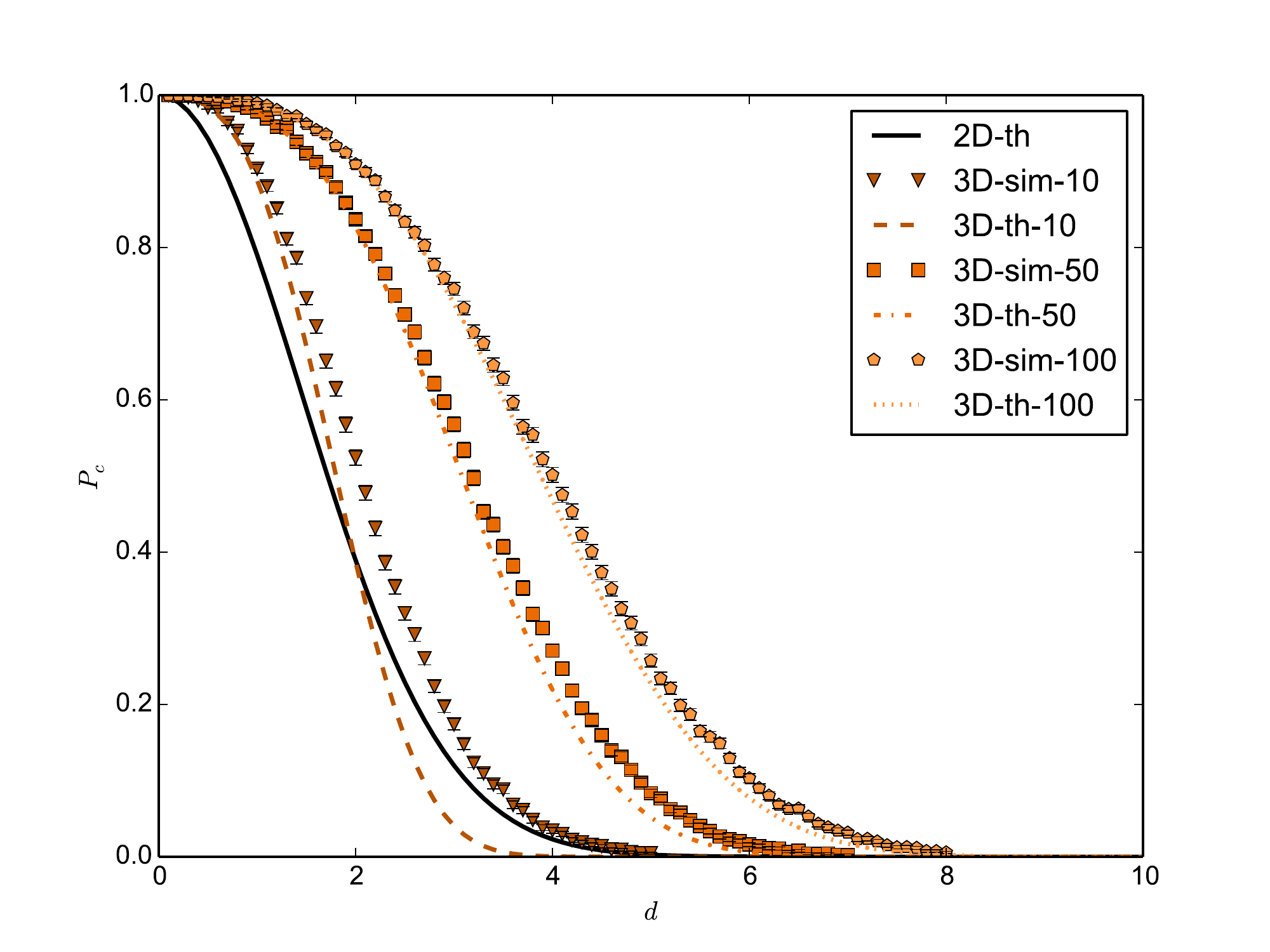}
  \caption{Comparison of theoretical and simulation results for PPP interference}
  \label{simpppfig}
\end{figure}

Fig. \ref{simpppfig} depicts the comparison of the probability of coverage $P_c$ between simulation and theoretical results for different building heights (parameter $Z$) in function of the emitter-receiver distance. For each simulation point, 10 000 PPP realizations are produced. The 95\% confidence interval is plotted, but is barely visible (it is very small). The 2D model corresponds to the solid curve, we observe that it underestimate the probability of coverage compared to 3D models and simulations. The three dashed and dotted curves correspond to the 3D theoretical models. They are obtained from expression (\ref{pc3Dpppeq}) with $\rho=\frac{\lambda}{Z}$ and $Z\in {10m,50m,100m}$. We observe, as expected, that the probability of coverage is higher when the considered height $Z$ is greater because the nodes are more spread and thus the distances to the interferers are larger on average. For $Z=50m$ and $Z=100m$ the simulations match the theoretical prediction very well. Nevertheless a tiny deviation can be noticed at the tail of the curves. This deviation is larger for $Z=50m$ than for $Z=100m$. In the $Z=10m$ case, the deviation of the theoretical model from the simulations is very significant. We can interpret these deviations as a border effect when the height of the box is reduced: the 3D theoretical model assumes no limits on the $x$, $y$ and $z$ axis. We also observe that even when the height of the box is reduced to $Z=10m$ the 2D theoretical model is not a good representation of the simulation data either. In this case, we note that the 3D model is a better match for the start of the curve and the 2D model is better for the tail of the curve.

We conclude that the 3D model seems to be accurate enough when the box height is sufficiently large to avoid border effect. Moreover, even when the height is small (the $Z=10m$ case) 3D model still provides a good representation of the probability of coverage close to the emitter because the border effect is larger for longer emitter-receiver distances.

In the next part of this paper, we consider another medium access model: CSMA. More precisely, we extend to $\mathbb{R}^3$ an approximation of the probability of coverage for the MMP model. We then compare the theoretical model predictions to simulation results.

\section{3D Modified Matern Process interference}
\label{MMPsec}

In this section, we treat the 3D case for the Modified Matern type II Process (MMP) which is considered in many works \cite{nguyen07,elsawy13} in order to model CSMA access. With this model, nodes too close to each others cannot transmit at the same time, interference is thus reduced. The MMP model is detailed in Section \ref{MMPmodelsec}.  

Since the exact Laplace functional is not known for the MMP, the technique used in the previous section for PPP (notably step (c) in equation (\ref{laplaceeq})) cannot be applied directly. In previous works, several different approximation techniques are used \cite{nguyen07,hasan07,kaynia11,elsawy13}. In this paper, we use the technique which consists in:
\begin{enumerate}
 \item assuming a finite region containing the potential contenders for a node;
 \item deriving the resulting MMP intensity $\rho_{csma}$;
 \item considering only dominant interferers;
 \item approximating the MMP with a PPP of intensity $\rho_{csma}$ outside of the emitter contention domain.
\end{enumerate}

This approximation technique is notably used in \cite{elsawy13} for an alternative version MMP. In our work, we derive expressions for the 2D and 3D cases, whereas \cite{elsawy13} is only for the 2D case and the alternative MMP. This technique allows to obtain an expression of the probability of coverage for the MMP, despite the fact that the exact expression is not known. It is detailed for the 3D case in the remainder of this section.

\subsection{The Modified Matern type II Process}
\label{MMPmodelsec}

In this part of the paper, we use a modified version of the Matern type II process \cite{nguyen07}. The classic Matern type II process is built from a marked PPP: each point of the PPP is marked with a real $m\in [0,1]$ and the PPP is thinned by retaining only the points which have the smallest mark in a ball of radius $D$ centered on themselves. Formally, $\Phi_m = \{ x_i\in \Phi | m_{i} < m_{j}, \forall x_j \in \Phi \cap \mathcal{B}(x_i,D)\setminus x_i \}$, with $\Phi$ the PPP in $\mathbb{R}^3$, $m_i$ the mark of point $x_i$ and $\mathcal{B}(x_i,D)$ a ball of radius $D$ centered on $x_i$. In the modified version of the process, there is no fixed radius $D$, but the radius is replaced by the notion of detection: a node from the marked PPP is kept in the MMP if it does not detect any signal from a node with a lower mark. The signal from a node is detected if it is above a threshold $T_d$. So formally the process is defined as $\Phi_m = \{x_i\in \Phi | m_{i} < m_{j},  \forall x_j : P_t h_{ij} r_{ij}^{-\alpha} \geq T_d  \}$, with $P_t$ the transmission power, $h_{ij}$ and $r_{ij}$ the fading coefficient and distance between $x_i$ and $x_j$ respectively.

The MMP allows to model the position of the nodes accessing the medium concurrently with a CSMA MAC scheme. The underlying PPP represents the positions of the nodes. In the following subsections, we detail the steps of the aforementioned approximation technique applied for the 3D case.

\subsection{Defining a finite region for contenders}

In reality, the region which contains the contenders of a given node (the nodes which can be detected by that node) presents an irregular shape because of the randomness of the fading coefficients. Nevertheless, as mentioned previously, the first step of the approximation technique consists in considering a fixed radius, called the detection radius (denoted $r_d$), outside of which the probability (denoted $\epsilon_d$) that a node is detected is arbitrary low. Thus the contenders of a node $x_i$ are contained in $\mathcal{B}(x_i,r_d)$ which is a ball in the 3D case. It allows to simplify the expressions without losing too much accuracy \cite{elsawy13}. Formally, the detection radius is defined as follows:
\begin{align}
 P(P_th_{ij}r_d^{-\alpha}\geq T_d)\leq \epsilon_d & \Leftrightarrow r_d = \left(\frac{P_t}{T_d} F_{H_{ij}}^{-1}(\epsilon_d)\right)^{1/\alpha} \nonumber\\
 & \Leftrightarrow r_d = \left(\frac{P_t}{T_d} \frac{-\ln(\epsilon_d)}{\mu} \right)^{1/\alpha}
\end{align}

if we consider $H_{ij}\sim exp(\mu)$.

\subsection{The intensity of the Modified Matern Process}
\label{intensitysec}

The intensity of the resulting process (considering only contenders in $\mathcal{B}(x_i,r_d)$) is given as $\rho_{csma}=\rho P_{csma}$ \cite{stoyan87}, with $P_{csma}$ the probability for a node of the underlying PPP to be retained and $\rho$ the intensity of the PPP. Thus $P_{csma}$ is the probability that among the contenders of a node $x_i$ (the nodes of $\Phi$ which are inside $\mathcal{B}(x_i,r_d)$ and that are detected by $x_i$) none have a smaller mark than $x_i$.

Formally, it is given as:

\begin{align}
 \label{pcsmaeq}
 P_{csma}&=\sum \limits_{k=0}^{+\infty} \sum \limits_{n=k}^{+\infty} \frac{1}{1+k} P_n \binom{n}{k} P_d^k (1-P_d)^{n-k} \nonumber\\
 &=\frac{1-e^{-\rho (4/3) \pi r_d^3 P_d}}{\rho (4/3) \pi r_d^3 P_d}
\end{align}

The details of the argument are similar as those of \cite{nguyen07} or \cite{elsawy13} and thus they are not reproduced here. $P_n$ and $P_d$ are respectively the probability to have $n$ nodes in $\mathcal{B}(x_i,r_d)$ and the probability to detect a node which is in $\mathcal{B}(x_i,r_d)$. $P_n$ is thus the probability that there are $n$ nodes in a volume of $(4/3) \pi r_d^3$ which is given by the Poisson distribution. For $P_d$, in the 3D case we have:
\begin{align}
 P_d&=P(P_t h_{ij} r_{ij}^{-\alpha}\geq T_d) \nonumber\\
 & = \mathbb{E}_{R_{ij}}\left[ e^{-\mu T_d r_{ij}^{\alpha} / P_t} \right] \nonumber\\
 & \stackrel{(f)}{=} \int \limits_{0}^{r_d} \frac{3 r_{ij}^2}{r_d^3} e^{-\mu T_d r_{ij}^{\alpha} / P_t} dr_{ij} \nonumber\\
 & = \frac{3}{r_d^3} \frac{\Gamma(\frac{3}{\alpha})-\Gamma(\frac{3}{\alpha},\frac{ \mu T_dr_d^{\alpha}}{P_t })}{\alpha \left(\frac{\mu T_d}{P_t}\right)^{3/\alpha}} \label{pdeq}
\end{align}
with $\Gamma(a)$ and $\Gamma(a,b)$ the Gamma function and incomplete upper Gamma function, (f) follows from  $H_{ij}\sim exp(\mu)$ and $f_{R_{ij}}(r_{ij})$ is given by:
\begin{equation}
  f_{R_{ij}}(r_{ij})=\frac{3 r_{ij}^2}{r_d^3}
  \label{frijeq}
\end{equation}

$f_{R_{ij}}(r_{ij})$ is the density function of the random distance between $x_i$ and $x_j$ in $\mathcal{B}(x_i,r_d)$. In order to prove that expression (\ref{frijeq}) is correct for the 3D case (for the 2D case it is given in \cite{sousa90}) we have to remember that $x_j$ is uniformly placed in $\mathcal{B}(x_i,r_d)$ because of the underlying PPP properties, so we have $F_{R_{ij}}(r_{ij})=\frac{(4/3)\pi r_{ij}^3}{(4/3)\pi r_d^3}$. Taking the derivative with respect to $r_{ij}$ yields the result. Expression (\ref{pdeq}) differs from the 2D case where $f_{R_{ij}}(r_{ij})$ is $\frac{2 r_{ij}}{r_d^2}$ \cite{sousa90}, but the expression remains reasonably easy to compute in the 3D case (the Gamma functions can be efficiently computed).

From (\ref{pcsmaeq}) and $\rho_{csma}=\rho P_{csma}$, we conclude that the intensity of the MMP is:

\begin{equation}
 \rho_{csma}=\frac{1-e^{-\rho (4/3) \pi r_d^3 P_d}}{(4/3) \pi r_d^3 P_d}
 \label{rcsmaeq}
\end{equation}

\subsection{Dominant interferers vulnerability radius}

Once we have obtained the intensity of the MMP, we consider only the interferers which can corrupt the signal on their own and we assume that they all lie in a region around the receiver, called the vulnerability region. Similarly to the detection radius, we define the vulnerability radius as the radius for which nodes beyond that limit have an arbitrary low probability (again noted $\epsilon_v$) to make the SIR at the receiver drop under the reception threshold. The receiver is placed at the origin and noted $o$. The test emitter is part of the process and noted $x_i$. Formally the vulnerability radius is defined as follows:

\begin{align}
 P\left( \frac{h_{io} r_{io}^{-\alpha}}{h_{jo} r_{v}^{-\alpha}}\leq \beta\right)\leq \epsilon_v & \Leftrightarrow \epsilon_v = F_{\frac{H_{io}}{H_{jo}}}\left(\beta \frac{r_v^{-\alpha}}{r_{io}^{-\alpha}} \right)\nonumber\\
 & \Leftrightarrow r_v=r_{io}\left(\beta \frac{1- \epsilon_v}{\epsilon_v} \right)^{1/\alpha}
\end{align}

with $\frac{H_{io}}{H_{jo}}$ the ratio of exponential random variables of parameter $\mu$ so $F_{\frac{H_{io}}{H_{jo}}}(l)=1-\frac{1}{1+l}$.

\subsection{Probability of coverage}

We give an approximation of the probability of coverage ($P_{c}^{csma}$) for a typical receiver at the origin. As previously mentioned, the MMP is approximated with a PPP of the same intensity outside of the contention domain of the emitter $x_i$, this approximation is known to be very close to simulation results \cite{haenggi11}. In this context, the probability of outage is the probability that among the nodes in the vulnerability radius which coexist with $x_i$ (outside of the contention region of $x_i$) some will be able to make the SIR drop under the threshold $\beta$. The probability of coverage is the complement of the probability of outage:

\begin{align}
 \label{pcmmpeq}
 P_{c}^{csma}& =1-\sum \limits_{n=1}^{+\infty} \sum \limits_{k=n}^{+\infty} \sum \limits_{t=1}^{n} P_k \binom{k}{n} (1-P_{d'})^n P_{d'}^{k-n} \binom{n}{t} P_{\beta}^t (1-P_{\beta})^{n-t} \nonumber\\
 & = e^{-K_{csma}P_{\beta}(1-P_{d'})}
\end{align}

with $P_k=\frac{(K_{csma})^k e^{-K_{csma}}}{k!}$, $K_{csma}=\rho_{csma}(4/3)\pi r^3_v$, $P_{d'}$ and $P_{\beta}$ are respectively the probability that a node coexists with the emitter and the probability that the SIR drops under the reception threshold. $P_{d'}$ and $P_{\beta}$ are derived below in this section. The full detail argument for a result close to Equation (\ref{pcmmpeq}) (2D case) can be found in \cite{elsawy13} (equation (\ref{pcmmpeq}) derivation only involves well-known convergent series). Deep differences from the 2D case are found in the expressions of $P_{d'}$ and $P_{\beta}$ derived below (and also in expression (\ref{rcsmaeq}) of $\rho_{csma}$ as shown in Section \ref{intensitysec}). 

$P_{d'}$ is the probability for a node in $\mathcal{B}(o,r_v)$ to be in the contention domain of $x_i$ (to be detected by $x_i$):

\begin{align}
 \label{pdprimeq}
 P_{d'}&=P(P_t h_{ij} r_{ij}^{-\alpha}\geq T_d)\nonumber \\
 & = \mathbb{E}_{R_{ij}}\left[P\left(h_{ij} \geq \frac{T_dr_{ij}^{\alpha}}{P_t } | r_{ij} \right)\right]\nonumber \\
 & = \int \limits_{0}^{r_v + r_{io}} f_{R_{ij}}(r_{ij}) e^{-\mu T_d r_{ij}^{\alpha} / P_t} dr_{ij}\nonumber \\
 & = \frac{3 \Gamma\left( \frac{3}{\alpha}\right) - 3 \Gamma\left(\frac{3}{\alpha},A\right)}{\alpha\,{r_v}^{3}\,{\left( \frac{\mu\,Td}{Pt}\right) }^{\frac{3}{\alpha}}}\nonumber\\
 & + \left[3C(-r_v^4-2r_{io}r_v^3+2 r_{io}^3r_v+r_{io}^4)\right. A^{(4/\alpha)} B^{(2/\alpha)}\nonumber \\
 & -6D( r_{io}r_v^3+3r_{io}^2r_v^2+3r_{io}^3 r_v+r_{io}^4) A^{(4/\alpha)}B^{(1/\alpha)}\nonumber \\
 & +3E(r_v^4+4r_{io}r_v^3+6r_{io}^2r_v^2+4 r_{io}^3r_v+r_{io}^4) A^{(4/\alpha)}\nonumber \\
 & +B^{(4/\alpha)}\left\{3F(r_v^4-2r_{io}r_v^3+2r_{io}^3r_v-r_{io}^4)\right.A^{(2/\alpha)}\nonumber \\
 &+6G(r_{io}r_v^3-3r_{io}^2r_v^2+3 r_{io}^3r_v-r_{io}^4)A^{(1/\alpha)}\nonumber \\
 &\left. \left. +3H(- r_v^4+4r_{io}r_v^3-6r_{io}^2r_v^2+ 4r_{io}^3r_v-r_{io}^4)\right\}\right] \nonumber \\
 & / \left[4\alpha r_{io}r_v^3A^{(4/\alpha)} B^{(4/\alpha)}\right]
\end{align}

with
$A=\left(\frac{\mu\,{\left( r_v-r_{io}\right) }^{\alpha}\,Td}{Pt}\right)$,
$B=\left(\frac{\mu\,{\left( r_v+r_{io}\right) }^{\alpha}\,Td}{Pt}\right)$,
$C=\Gamma(2/\alpha,B)$,
$D=\Gamma(3/\alpha,B)$,
$E=\Gamma(4/\alpha,B)$,
$F=\Gamma(2/\alpha,A)$,
$G=\Gamma(3/\alpha,A)$ and
$H=\Gamma(4/\alpha,A)$.

\begin{figure*}[ht]
        \centering
        \subfloat[$0\leq r_{ij} \leq r_v - r_{io}$]{
                \includegraphics[width=2.1in, keepaspectratio=true]{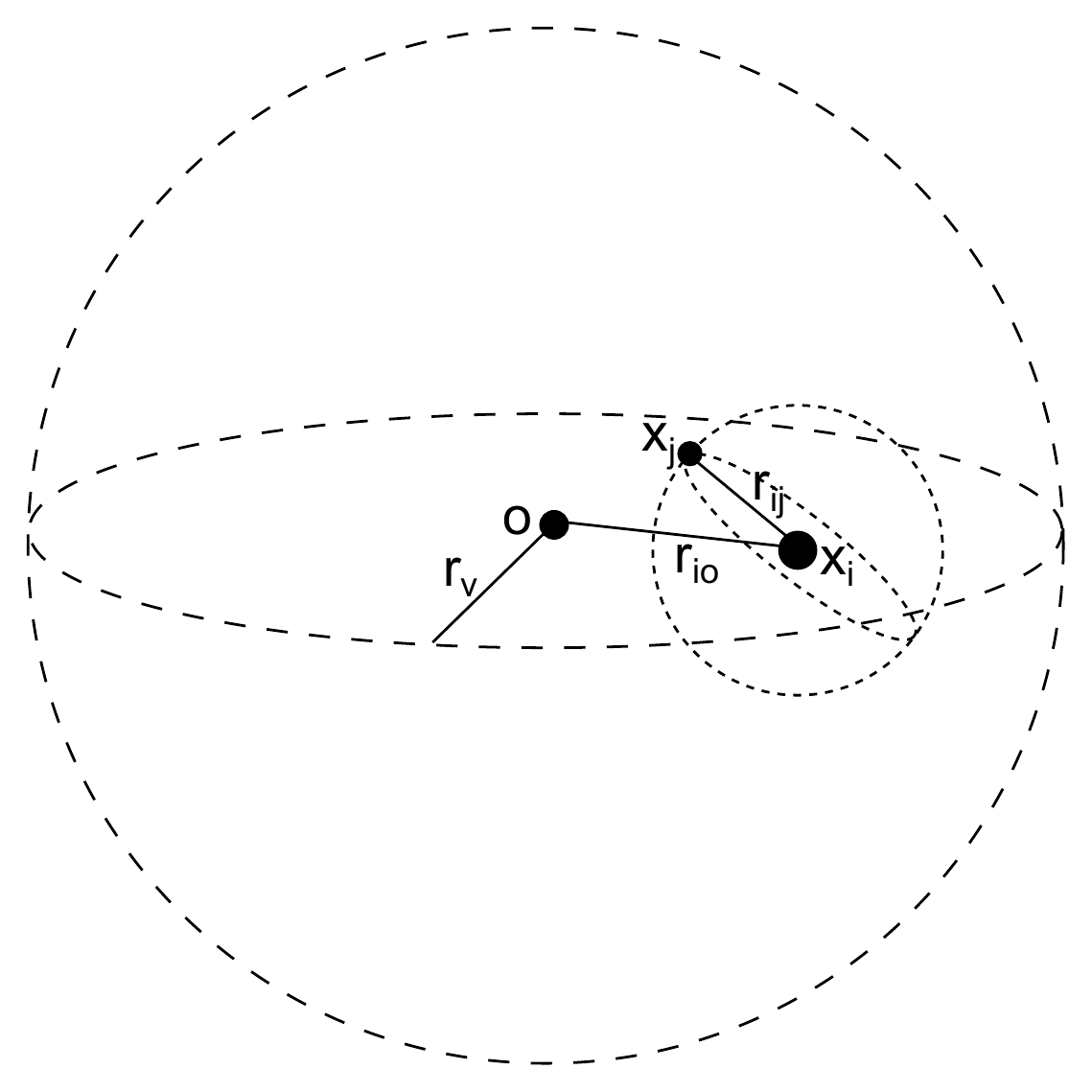}
                \label{outfig}
        }
        \subfloat[$r_v - r_{io} < r_{ij} \leq r_v+r_{io}$]{
                \includegraphics[width=2.5in, keepaspectratio=true]{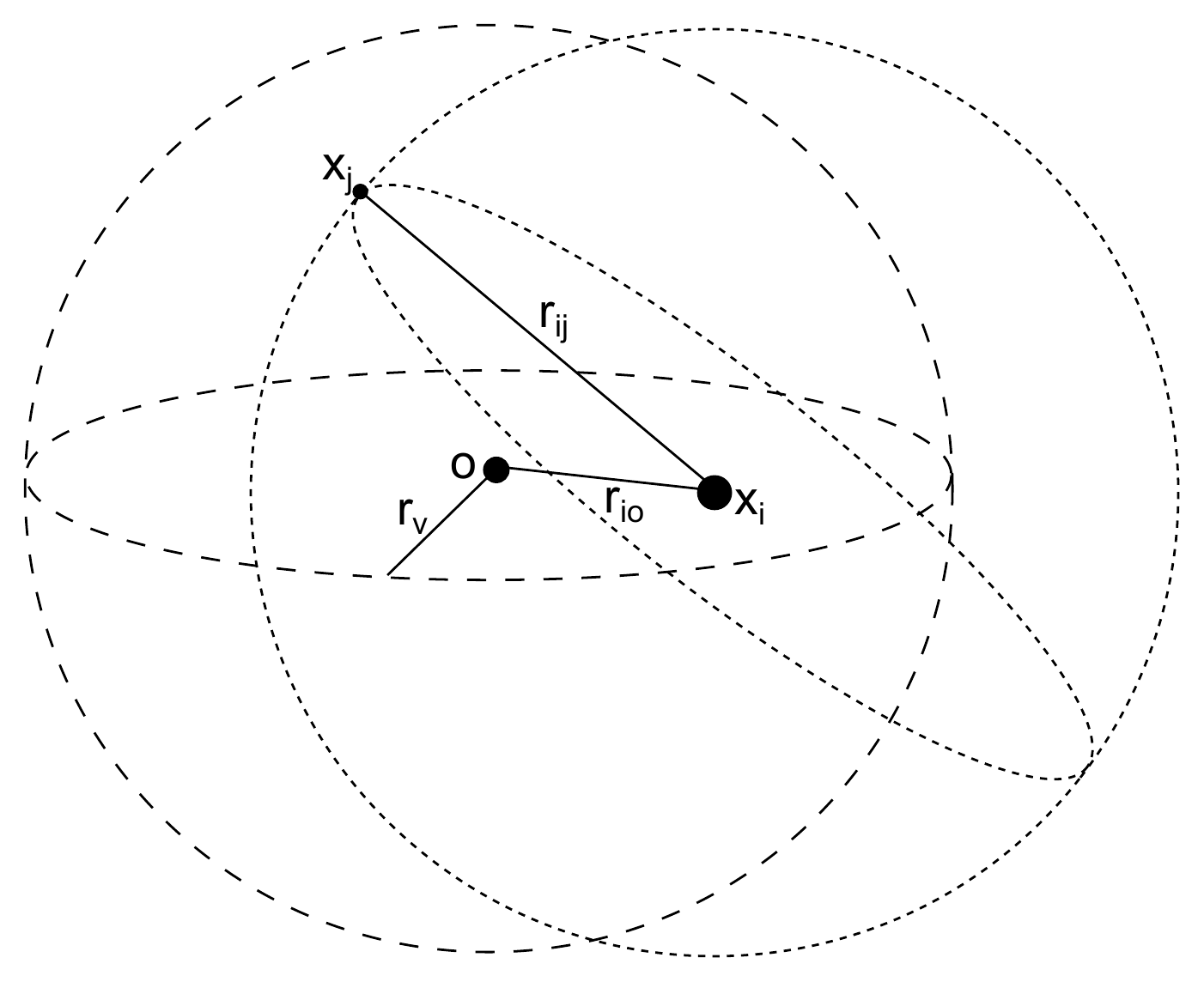}
                \label{infig}
        }
        \caption{Representation of the distances $r_{ij}$, $r_{io}$ in the vulnerability radius}
        \label{spherefig}
       
\end{figure*}

Here the expression of $f_{R_{ij}}(r_{ij})$ changes from (\ref{frijeq}) in Section \ref{intensitysec} because we are no more interested in $\mathcal{B}(x_i,r_d)$ but rather in $\mathcal{B}(o,r_v)$. As can be seen in Fig. \ref{spherefig} we now have two cases: either $\mathcal{B}(x_i,r_{ij})$, the dotted sphere, is contained in $\mathcal{B}(o,r_v)$, the dashed sphere ($0\leq r_{ij} \leq r_v - r_{io}$), or it is not ($r_v - r_{io} < r_{ij} \leq r_v+r_{io}$). In the first case, the expression is similar to (\ref{frijeq}). In the second case, the expression for $f_{R_{ij}}(r_{ij})$ changes because we know that $x_j$ cannot be outside of $\mathcal{B}(o,r_v)$ so it must lie in the intersection of $\mathcal{B}(x_i,r_{ij})$ and $\mathcal{B}(o,r_v)$:

\begin{equation}
   f_{R_{ij}}(r_{ij}) 
   \begin{cases} 
      \frac{3 r_{ij}^2}{r_v^3}, \text{ for } 0\leq r_{ij} \leq r_v - r_{io} \\
      \frac{3r_{ij}\left( r_v-r_{io}+r_{ij}\right) \left( r_v+r_{io}-r_{ij}\right) }{4 r_{io} r_v^3},\text{ for } r_v - r_{io} < r_{ij} \leq r_v+r_{io}\\
      
   \end{cases}
\end{equation}

Interestingly, we note that the second piece of the expression of $f_{R_{ij}}(r_{ij})$ is rational whereas in the 2D case it includes trigonometric functions (arcsine) \cite{elsawy13}. It allows to have an exact expression for $P_{d'}$ in the 3D case (expression (\ref{pdprimeq})) whereas, to the best of our knowledge, numerical integration has to be used in the 2D case.

$P_{\beta}$ is the probability that a node in $\mathcal{B}(o,r_v)$ is able to make the SIR at the receiver drop under the reception threshold $\beta$:

\begin{align}
 P_{\beta}&=P\left( \frac{h_{io} r_{io}^{-\alpha}}{h_{jo} r_{jo}^{-\alpha}}\leq \beta\right) \nonumber\\
 &=\mathbb{E}_{R_{jo}}\left[P\left(\frac{h_{io}}{h_{jo}} \leq \beta r_{jo}^{-\alpha} r_{io}^{\alpha} \right)\right] \nonumber\\
 &=\int \limits_{0}^{r_v} \frac{3 r_{jo}^2}{r_v^3} F_{\frac{H_{io}}{H_{jo}}}( \beta r_{jo}^{-\alpha} r_{io}^{\alpha}) dr_{jo} \nonumber\\
 &\stackrel{(g)}{=} \int \limits_{0}^{r_v} \frac{3 r_{jo}^2}{r_v^3}  \left(\frac{\beta}{\beta+\left(\frac{r_{jo}}{r_{io}}\right)^{\alpha}} \right)dr_{jo} \nonumber\\
 &\stackrel{(h)}{=} \frac{r_{io}^3}{r_v^3} \beta^{3/\alpha} \int \limits_{0}^{\left(\frac{r_v}{r_{io}\beta^{1/\alpha}}\right)^3} \frac{1}{1+w^{\alpha/3}} dw \nonumber\\
 & = {}_2F_1(1,\frac{3}{a},\frac{3}{a}+1,-{\left(\frac{r_v}{r_{io}\beta^{1/\alpha}}\right)}^{a})
\end{align}
where ${}_2F_1$ is the Gaussian hyper-geometric function, (g) follows from $F_{\frac{H_{io}}{H_{jo}}}(l)=1-\frac{1}{1+l}$ and $f_{R_{jo}}(r_{jo})=\frac{3 r_{jo}^2}{r_v^3}$, and (h) from the change of variable $w=\left(\frac{r_{jo}}{r_{io}\beta^{1/\alpha}}\right)^3$.

\begin{figure}[ht]
  \centering
  \includegraphics[width=3in, keepaspectratio=true]{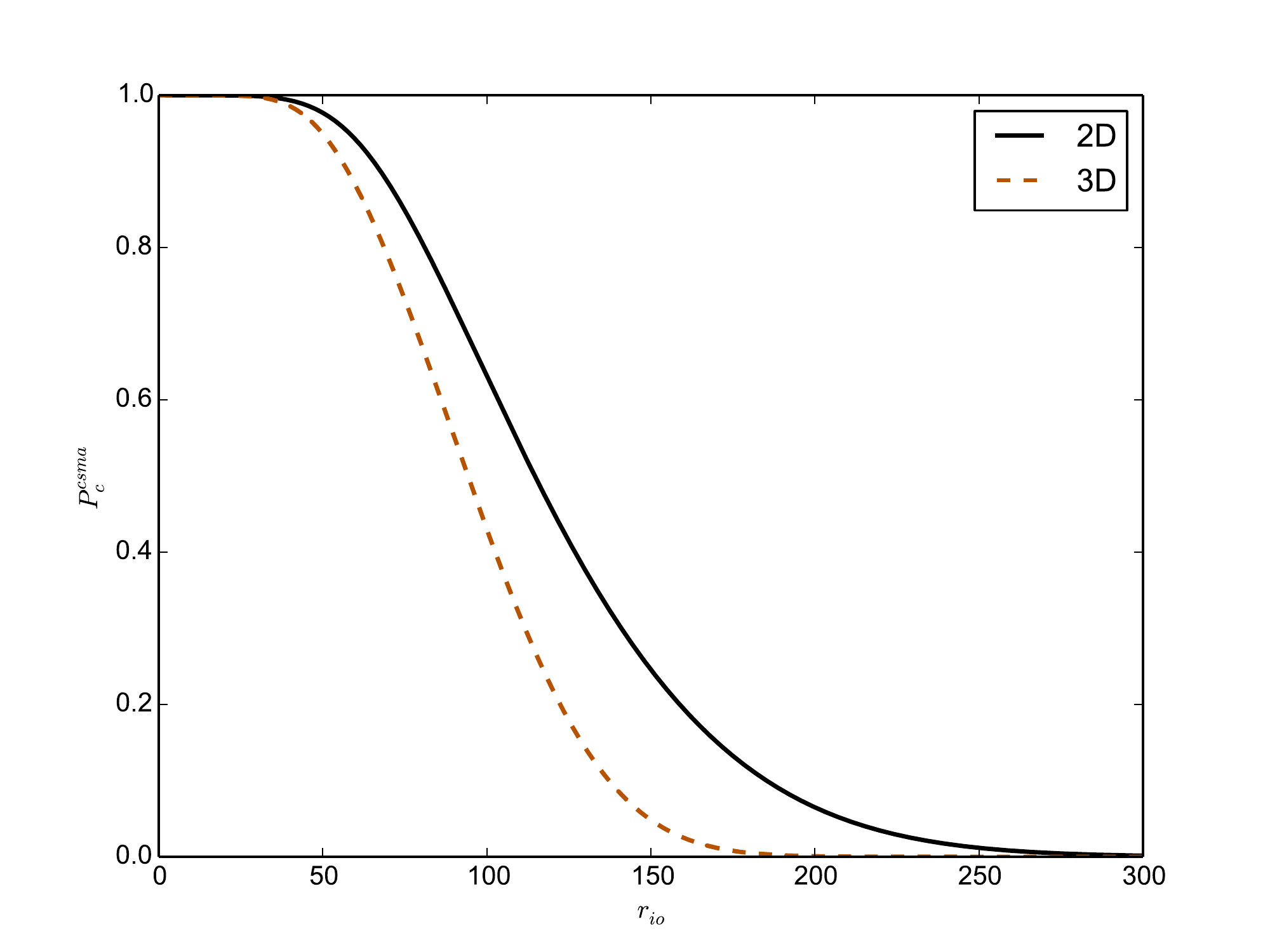}
  \caption{Probability of coverage for 2D and 3D MMP}
  \label{csmafig}
\end{figure}

Fig. \ref{csmafig} depicts the probability of coverage for the 2D and 3D cases for the MMP. $P_c^{csma}$ (Equation (\ref{pcmmpeq})) is given in function of $r_{io}$, the emitter-receiver distance. The curves are plotted with the values $P_t=100$mW, $T_d=-76$dBm, $\beta=10$, $\alpha=4$, $\mu=1$, $\epsilon_d=10^{-6}$, $\epsilon_v=10^{-2}$, and $\rho$ and $\lambda$ from Section \ref{introsec}. First we have to note that in this case, the communication range is more realistic (even if in reality in dense urban environment, we may encounter pathloss exponents greater than $4$). We observe that, with the aforementioned parameter values, the probability of coverage is strictly higher in the 2D case for emitter-receiver distances above $50m$. We can interpret this result as follows: for the PPP in Section \ref{PPPsec}, we remarked that the 3D probability of coverage tends to drop later than the 2D, but then the slope is steeper. For the MMP model, the difference in the drop position is suppressed by the contention mechanism (it avoids close interferers), but the difference in the slopes remains.

\begin{figure}[ht]
  \centering
  \includegraphics[width=3in, keepaspectratio=true]{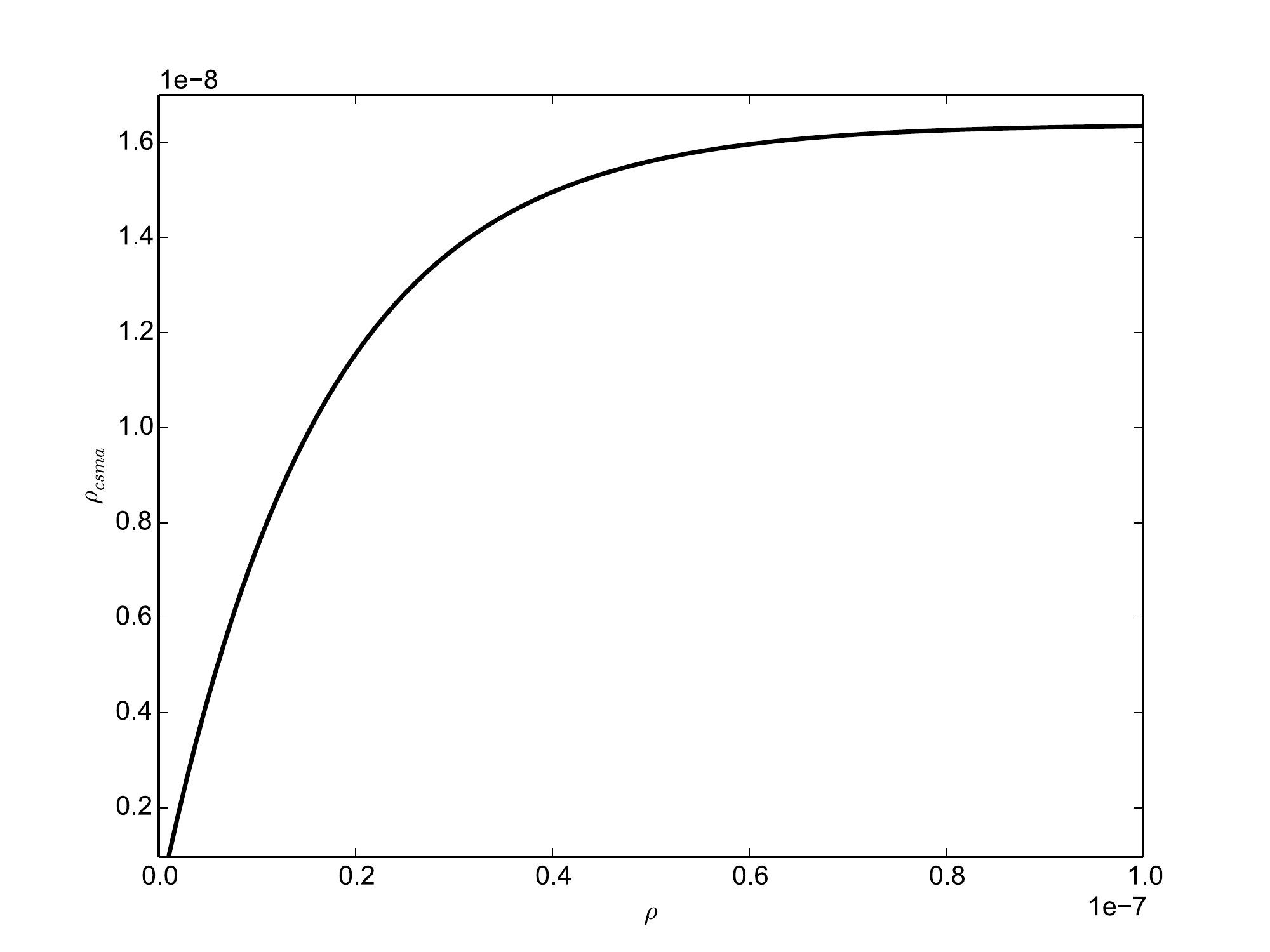}
  \caption{Intensity of the MMP in function of the intensity of the underlying PPP}
  \label{rcsmafig}
\end{figure}

As described in Section \ref{intensitysec}, the intensity $\rho_{csma}$ of the point process representing the transmitters (the MMP) is different from the underlying PPP intensity $\rho$. Fig. \ref{rcsmafig} depicts the value of $\rho_{csma}$ in function of $\rho$ (it corresponds to equation (\ref{rcsmaeq})), we observe that the value of $\rho_{csma}$ converges toward a maximum value: $\lim\limits_{\rho \to \infty} \rho_{csma} = \frac{1}{(4/3) \pi r_d^3 P_d} $. It means that, when the intensity of nodes increases, the intensity of interferers converges. In the remainder of the paper, we call this situation saturated condition. The convergence implies that even in the 2D case where all the nodes are projected on the Euclidean plane, the intensity of interferers will not grow as much as in the PPP case.

\subsection{Comparison with simulation results}
\label{mmpsimsec}

In this section, we compare the 3D theoretical model with the 2D theoretical model and with simulation results. The goal, as in Section \ref{pppsimsec}, is to evaluate the adequacy of 2D and 3D models to represent the probability of coverage for different node parameters and building heights. In this section, we experiment with two simulation setups: the first with 802.11 node parameters, and the second with node parameters representing 802.15.4 radios.

The first simulation setup is as follows: the underlying PPP is generated in box of $2000m\times 2000m \times Z$ with $Z\in {20m,200m,2000m}$. We use a larger box than for the PPP case because otherwise we do not have enough nodes in the MMP after the thinning of the PPP through the contention process. The contention process consists in nodes picking a random mark uniformly in $[0,1]$. Then each node checks if it can detect neighbors with a lower mark. A node is detected if the signal received is above a threshold (the signal strength depends on the transmission power, the distance, and on a fading random variable). If no detectable neighbor has a lower mark, the node is kept in the MMP.

As in Section \ref{pppsimsec}, the 3D intensity of the process is calculated based on the 2D intensity: $\rho=\frac{\lambda}{Z}$. In this case, we take $\lambda=7\times 10^{-5}$ so that we have enough nodes in the box and the thinning process is not too long (it is notably very long for the parameters considered in Section \ref{introsec}). We will see that this value is important only if we are not in saturated condition (see the comment on Fig. \ref{rcsmafig} in the previous section). For each realization of the MMP, we select the closest node to the center of the box to be the emitter, we than compute the SIR at $r_{io}$ meters from the emitter and check if it is over or below the reception threshold. We repeat this process $1000$ times for every $d$. The parameters of the node and the channel are as follows: $P_t=100$mW, $T_d=-76$dBm, $\beta=10$, $\alpha=4$, $\mu=1$, $\epsilon_d=10^{-6}$ and $\epsilon_v=10^{-2}$.

\begin{figure}[ht]
  \centering
  \includegraphics[width=3in, keepaspectratio=true]{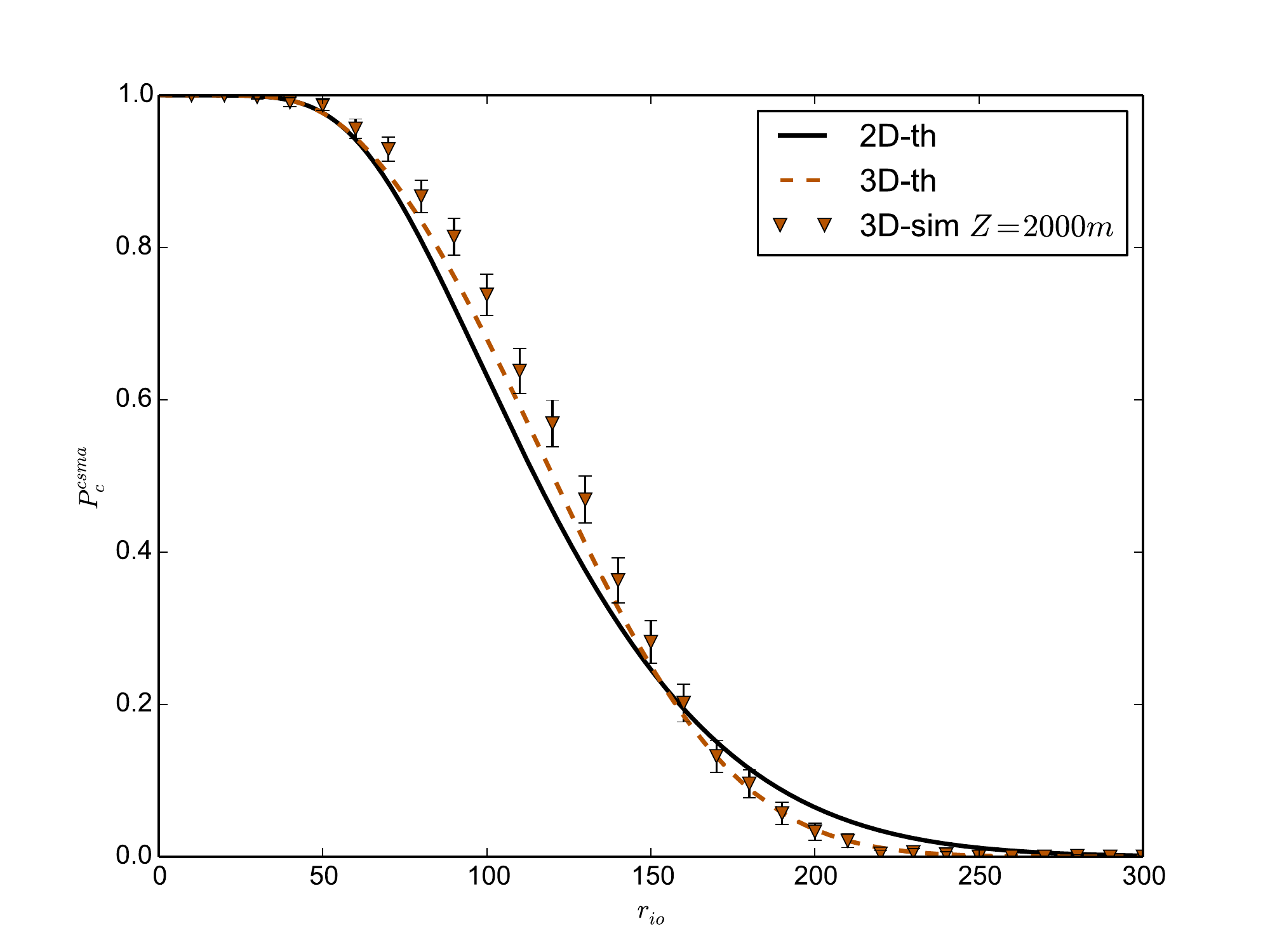}
  \caption{Comparison of theoretical and simulation results for the MMP model for $Z=2000m$}
  \label{sim2000fig}
\end{figure}

Fig. \ref{sim2000fig} depicts the simulation results  with $Z=2000m$ and 2D (solid curve) and 3D (dashed curved) model predictions for the probability of coverage. In this case, the 3D model is not in saturated condition. We observe that the 2D and 3D models yield very close values. Nevertheless, the simulation results closely follows the 3D model predictions.

\begin{figure*}[ht]
        \centering
        \subfloat[$Z=200m$]{
                \includegraphics[width=3in, keepaspectratio=true]{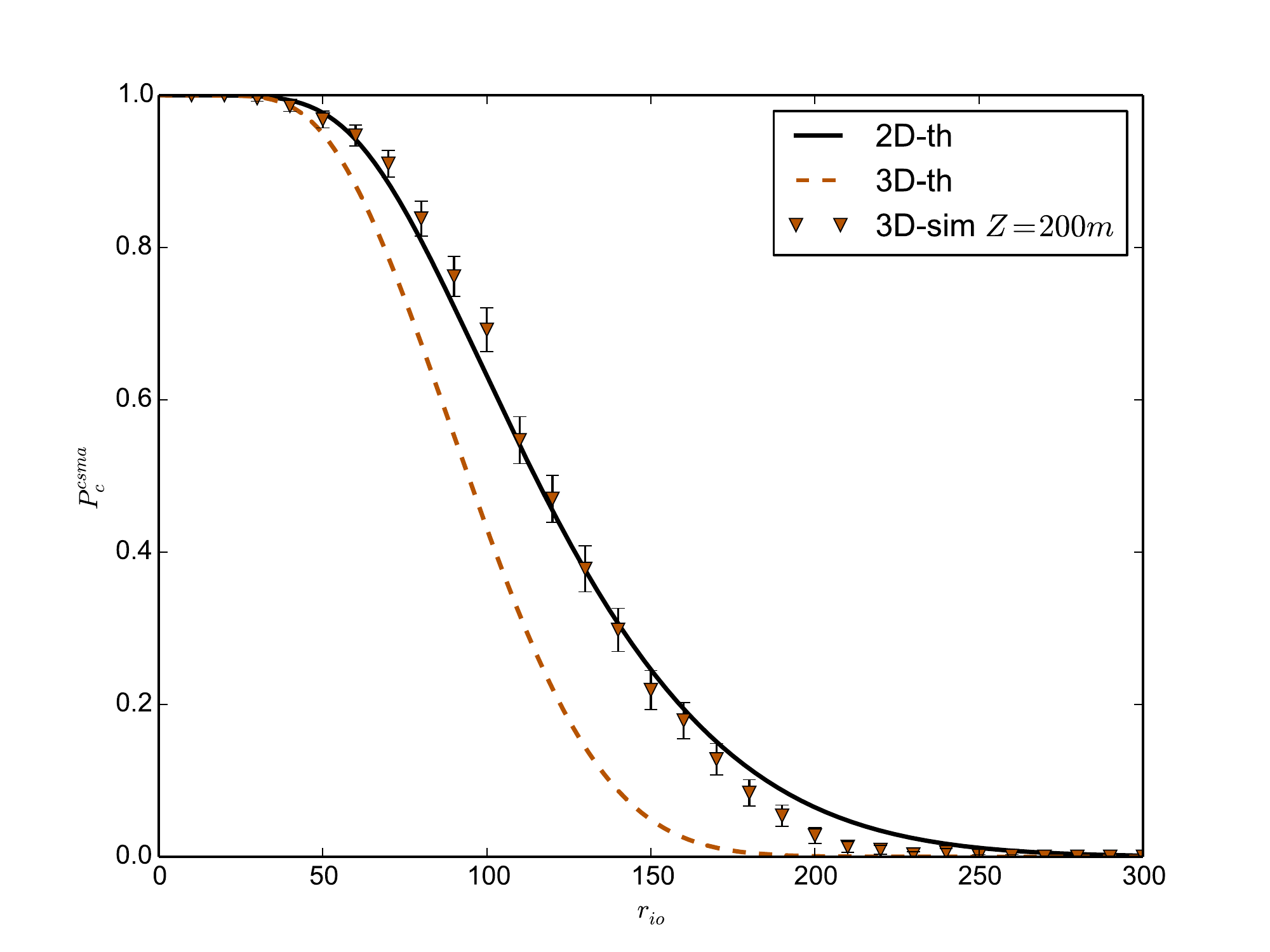}
                \label{sim200fig}
        }
        \subfloat[$Z=20m$]{
                \includegraphics[width=3in, keepaspectratio=true]{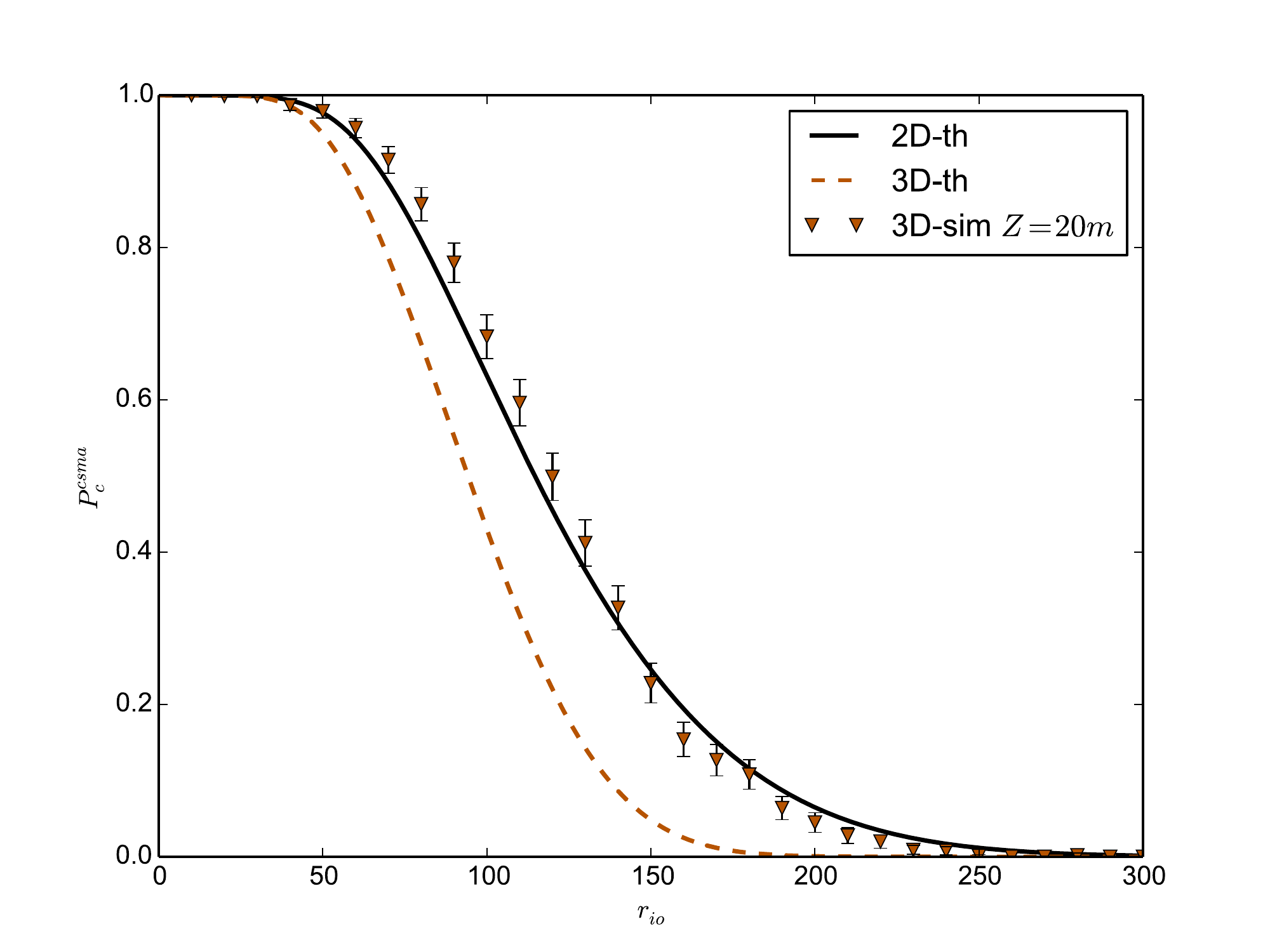}
                \label{sim20fig}
        }

        \caption{Comparison of theoretical and simulation results for the MMP model}
\end{figure*}

Figs. \ref{sim200fig} and \ref{sim20fig} present the same comparison as Fig. \ref{sim2000fig} but for more reasonable simulated $Z$: $200m$ and $20m$. In these cases, the 3D model is in saturated condition and we observe as in Fig. \ref{csmafig}, that the theoretical predictions of the 2D and 3D models differ. Interestingly, for both $Z=200m$ and $Z=20m$ the simulation results are closer to the 2D theoretical predictions. We note that for $Z=20m$ the match with the 2D model is better than for $Z=200m$ notably for the tail of the curve. These results can be explained by the protection radius granted to the emitter thanks to the contention process. Indeed, with the considered parameters, the average detection range is $251m$. Interferers are thus very rarely situated above or below the receiver in this context ($Z=200m$ and $Z=20m$) and so the vertical distances do not have a significant impact on the probability of coverage. We believe this account for the better match of the 2D model.

The second simulation setup is similar to the first except for $Td=-60dBm$ and $Pt=1mW$ (realistic for a typical WSN low power radio \cite{cc1100}) and the simulation box size is reduced to $200m\times 200m \times Z$ with $Z\in {20m,50m,75m,100m,200m}$.

\begin{figure}[ht]
  \centering
  \includegraphics[width=3in, keepaspectratio=true]{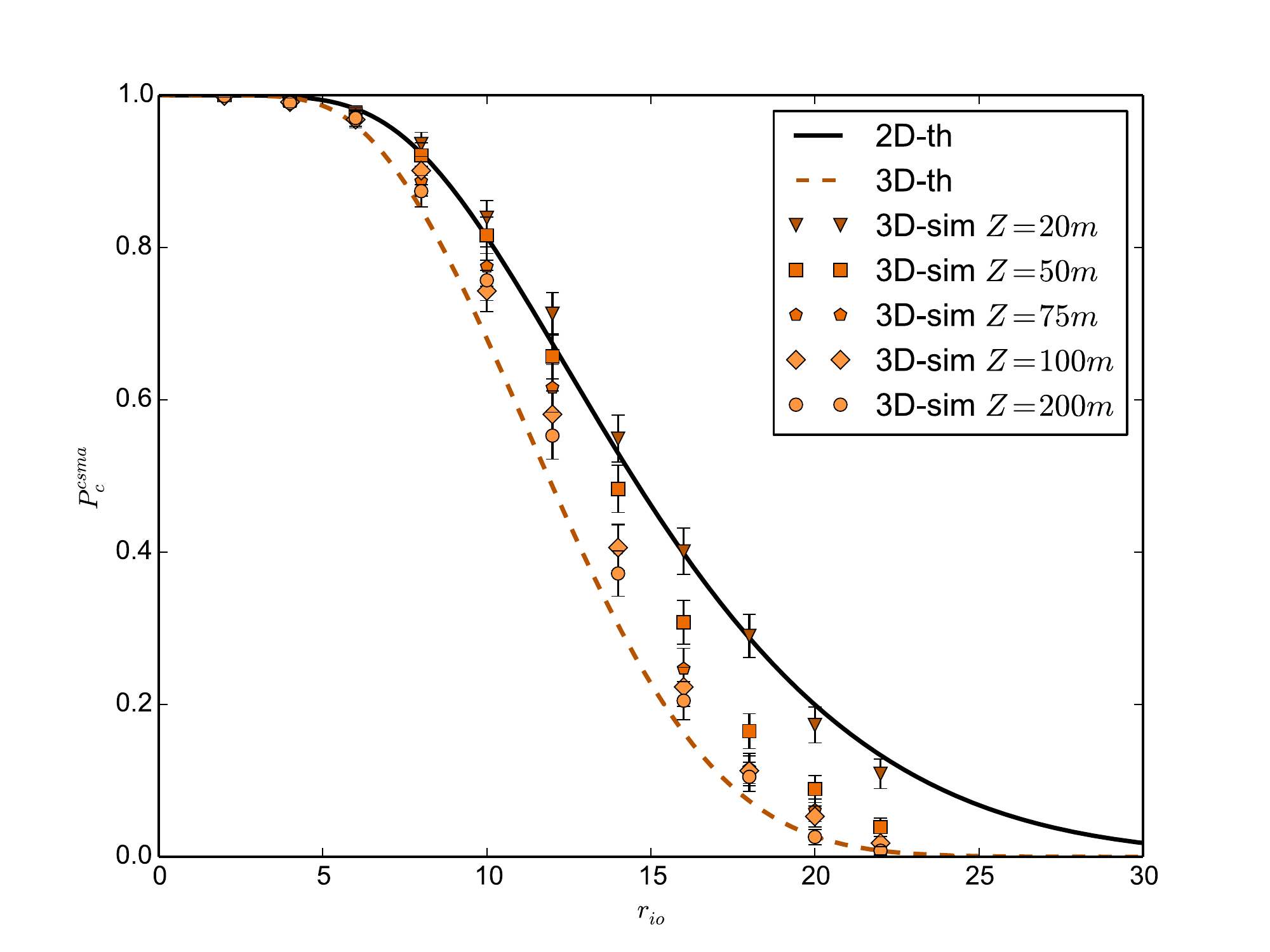}
  \caption{simu}
  \label{sim2Td60fig}
\end{figure}

Fig. \ref{sim2Td60fig} shows a comparison of the simulation results with 2D and 3D models for the second simulation setup. We are, again, in saturated conditions. In this case, we observe that the simulation results are distributed between the 2D and 3D theoretical models. When $Z$ increases the 3D model is a better match for the simulations results, while the 2D is better for smaller $Z$. In this case, the average detection range is $32m$. From the results presented in Fig. \ref{sim2Td60fig}, we conclude that the 3D model is a good representation of CSMA access in dense urban environment if the building height is at least two times greater than the average detection radius.

In this section, we highlighted the fact that in the CSMA case, the model has to be chosen carefully. Indeed, the 3D model seems to be relevant only if the average detection radius is smaller than the building height which is not the case for 802.11. Nevertheless, this model seems satisfying for low power low range communications for which the detection range is much smaller.

\section{Conclusion and future works}
\label{conclusec}

In this paper, we investigate the probability of coverage in the 3D case for two models: PPP and MMP. We show that abusively considering 2D networks when they are 3D can lead to either overestimating or underestimating the probability of coverage depending on the model parameters. We also notice interesting differences in the integral forms of the expressions when going from 2D to 3D which notably affect their tractability. 

By comparing the models with simulations, we observe that the 3D model is relevant in the PPP case even if the height of the simulation box is small compared to its length and width. In the MMP case for the modeling of CSMA, it depends on the detection radius. If the detection radius is larger than the building height, the 2D model is closer to the simulations, otherwise, the 3D model is more relevant.

Beside the results on stochastic geometry, this work can be seen as an incentive to use 3D model and 3D simulation layouts for the design and evaluation of protocol which is seldom done in reality. Section \ref{mmpsimsec} suggests that it is especially important for the design of low range IoT protocols in dense urban areas.

In the future, we plan to compare the model with traces collected in dense urban areas in order to validate our theoretical predictions and simulations. Moreover, it would be interesting to introduce a fourth dimension in the process in order to account for the packet arrivals which are spread in time.

\bibliographystyle{abbrv}
\bibliography{paper}   

\end{document}